\input harvmac
\Title{hep-th/9603089; FTUAM-96-11; PUPT-96-1601}
{ T-duality for open strings}
\smallskip
\centerline{Enrique Alvarez} 
\smallskip
\centerline{ Departamento de F\'{\i}sica Te\'orica, CXI}
\centerline{ Universidad Aut\'onoma, 28049 Madrid, Spain}
\centerline{\rm enrial@daniel.ft.uam.es}
\bigskip
\centerline{J.L.F.  Barb\'on}
\centerline{ Joseph Henry Laboratories}
\centerline{ Princeton University, Princeton, NJ, 08544 USA}
\centerline{\rm barbon@puhep1.princeton.edu}
\bigskip
\centerline{ and}
\bigskip
\centerline{J. Borlaf}
\centerline{ Departamento de F\'{\i}sica Te\'orica, CXI}
\centerline{ Universidad Aut\'onoma, 28049 Madrid, Spain}
\centerline{\rm javier@delta.ft.uam.es}
\baselineskip 18pt

\medskip

\noindent




\noblackbox
\parskip=1.5mm


\def\npb#1#2#3{{\it Nucl. Phys.} {\bf B#1} (#2) #3 }
\def\plb#1#2#3{{\it Phys. Lett.} {\bf B#1} (#2) #3 }

\def\prl#1#2#3{{\it Phys. Rev. Lett.} {\bf #1} (#2) #3 }
\def\mpla#1#2#3{{\it Mod. Phys. Lett.} {\bf A#1} (#2) #3 }


\def\dj{\hbox{d\kern-0.347em \vrule width 0.3em height 1.252ex depth
-1.21ex \kern 0.051em}}

\def\ket{\rangle}

\def\tX{\widetilde X}
\def\tx{\tilde x}
\def\tg{\tilde g}
\def\tb{\tilde b}
\def\tG{\tilde G}
\def\tB{\tilde B}
\def\tpsi{\tilde \psi} 

\def\bx{\overline x}
\def\bpsi{\overline \psi}
\def\btheta{\overline \theta}
\def\ba{\overline a}
\def\pt{\partial}
\def\zb{\bar z}


           \def\CO{{\cal O}} \def\CZ{{\cal Z}}
   
\def\CL{{\cal L}}   
  \def\CD{{\cal D}} 
 
\def\CJ{{\cal J}}



The T-duality transformations between open and closed superstrings in
different D-manifolds are generalized to curved backgrounds with
commuting isometries. We address some global aspects like the
occurrence  
of orientifold boundaries in general sigma models, 
 higher genus world sheets,  and the case of non-compact
isometries. The various world-volume effective actions are shown to
transform properly under T-duality. We also include a brief 
discussion of
the canonical transformations of boundary states in the operator
formalism.


\newsec{Introduction}
Target space T-duality is one of the most intriguing symmetries in
string theory. It was initially discovered as an invariance of toroidal
compactifications of closed strings under the change of the 
space time radius, $R$ to ${1\over R}$ \ref\j{K. Kikkawa and 
M. Yamasaki, \plb {149} {1984} {357 \semi}  
N. Sakai and I. Senda, {\it Prog. Theor. Phys.} {\bf 75} 
 (1976) 692.}, but it was 
soon realized that it is a much more generic property, and indeed a
systematic technique allows the study of this symmetry in all 
backgrounds with non-trivial group of isometries \ref\aagl{E.
 Alvarez, L. Alvarez-Gaum\'e and Y. Lozano, {\it Nucl. Phys.}
{\bf B41}
, Proc.  Suppl.
 (1995)1.}\ref\gpr{A. Giveon, M. Porrati, E. Rabinovici, {\it Phys. 
Rep. } 
 {\bf 244} (1994) 77.}\ref\buscher{T. H.      Buscher, \plb {194}  
{1987} {51},   \plb {201} {1988} {466.}}  
\ref\rv{M. Roc{\v e}k and E. Verlinde,  \npb {373} {1992} {630.}}. 
\par
While most earlier work dealt with closed strings only, thought
at the time the most interesting ones from the physical
point of view, recent work on string-string duality (cf., for
example, \ref\pold{ J. Polchinski, hep-th/9511157.} \ref\polwi{  
J. Polchinski and E. Witten, hep-th/9510169.}),
shows that there is a hidden unity between different types of 
string theories. In fact, already at a perturbative level,
 open strings can in a sense
be considered as closed strings in world-sheet orbifolds
\ref\wso{A. Sagnotti, in Non-Perturbative Quantum Field Theory,
eds. G. Mack et al. (Pergamon Press, 1988), 521
\semi M. Bianchi and A. Sagnotti, \plb  {247} {1990} {517;} 
\npb {361} {1991} {519 \semi}  P. Hor{\v a}va, \npb  
{327}{1989}{461;} \npb{418} {1994} {571;} \plb{289} {1992} {293.}}.  
The action of T-duality on open string theories has been
studied in the past \ref\Topen{J. Dai, R. G. Leigh and 
J. Polchinski, \mpla{4} {1989}  {2073 \semi} 
R. G. Leigh,  \mpla{4} {1989} {2767.}}  
\ref\greh{M. B. Green, \plb{266} {1991} {325\semi}
P. Hor{\v a}va, \plb{231}{1989}{251.}}. Recent excitement
followed Polchinski's idea that some exotic objects
appearing in the dual of Type I theory ({\it D-branes}) are the
 carriers of R-R charges needed for the implementation of the 
S-duality conjecture \ref\po{J. Polchinski,   \prl 
 {75} {1995} {4724.}}.
An excellent recent review can be found in  
  \ref\P{J. Polchinski, S. Chaudhuri and C.V.     Johnson, 
hep-th/9602052.}.
\par
The aim of the present paper is to examine some aspects of
these problems in more detail.  
The use of semiclassical methods allows  us to
 derive many of the well established facts about open string 
T-duality, in a simple and unified treatment and, most importantly,
provide an appropriate framework for   the
generalization to   
arbitrary spacetime backgrounds with (commuting)
isometries, even when an exact conformal field theory description is
not available. We also  pay   due attention to the effects
of non-orientable world-sheets in Type I theories, as well as
the subtleties of the T-duality/supersymmetry interplay
 \ref\aagb{E. Alvarez, L. Alvarez-Gaum\'e and I. Bakas,  
\npb{457}{1995}{3.}}.  

The paper is organized as follows. In section 2 we review the
auxiliary gauging procedure and apply it to the open string/D-brane
duality in the bosonic case,
 with a detailed discussion of the mapping of boundary
conditions, and the emergence of collective coordinates for the
D-brane in this formalism.
 In section 3 we discuss some new features arising in the
supersymmetric case, concerning the fermion boundary conditions. In
section 4 we check that the low energy effective world volume actions
(Dirac-Born-Infeld and Chern-Simons) have the desired covariance
properties under T-duality. Finally, we consider the higher genus case
in appendix A. Appendix B is a review of facts about T-duality for
non-compact isometries, and appendix C contains a very brief account  
of the canonical transformations involved in T-duality with
boundaries.


 \newsec {Open bosonic strings in background fields}
 Let us begin with the simplest bosonic model, which,
 while allowing for the most interesting physical phenomena,
 is devoid of complications due to supersymmetry.
We shall consider open and closed  bosonic strings propagating
     in an arbitrary $d$-dimensional 
 metric and (abelian) gauge field.
 Wess-Zumino antisymmetric tensors are not consistent if
 the theory is non-orientable, but we shall include
 them nevertheless for the time being. In modern language, we have a
closed string interacting with a Dirichlet  $(d-1)$-brane, and we
consider non-trivial massless backgrounds in the longitudinal
directions.   
\par
In the neutral
case (that is, the charge is opposite in both ends of 
the string),
the action can be written as\foot{We set $\alpha' =1$ throughout.}:
\eqn\uno{S = 
{1\over 4\pi
}\int_{\Sigma} (g_{\mu\nu}\eta^{ab} +i
\, b_{\mu\nu}\epsilon^{ab})\, \partial_a
x^{\mu}\partial_b x^{\nu} +
{i\over 2\pi}
\int_{\partial \Sigma} n_a A_{\mu}\partial_b x^{\mu} \epsilon^{ab}.}

The classification of allowed world-sheet
 topologies is much more complicated in the open case 
that in the more familiar closed one.
The Euler characteristic can be written, in a somewhat symbolic form, as
\eqn\cero{\chi = 2 -2g -c -b,}
where $g$ is the number of handles, $b$ the number
 of boundaries, and $c$ the number of crosscaps.
To the lowest order of string perturbation theory $\chi = 1$,
 only contribute the disc $D_2$ and in the non-orientable case
 the crosscap or, to be more precise, the two-dimensional real 
projective plane $P_2(R)$. (To the following ``one loop" order,
 corresponding to $\chi = 0$, we have  the annulus $A_2$,
 the M\"obius band $M_2$, 
and the Klein bottle $K_2 $). In this section we shall only consider
the leading contributions  from the disc and the crosscap.  
\par
The action \uno\  will be invariant under a target  isometry
with Killing vector $k^{\mu}$,  
 \eqn\dos{\delta_{\epsilon} x^{\mu} = \epsilon\,\, k^{\mu}(x)}
provided a vector $\omega_{\mu}$ 
and a scalar $\varphi$ exist, such that:
\eqn\tres{\eqalign{ \CL_k
\, g_{\mu\nu} =&0 \cr 
\CL_k \, b_{\mu\nu} =&
 \partial_{\mu}\omega_{\nu} - \partial_{\nu}\omega_{\mu} \cr  
\CL_k \, A_{\mu} =&  - 
\omega_{\mu} + \partial_{\mu} \varphi,}}
where $\CL_k$ represents the Lie derivative
 with respect to the Killing vector.

In the neutral case, it is clear that the boundary term representing 
the coupling of the background 
gauge field to the open string can be incorporated in the bulk
action  through the simple 
substitution  \foot{When the charges at both ends of the string
do not add to zero, one cannot get rid completely of the boundary 
term. We shall comment on a similar situation later in the main
 text.}
\eqn\seis{b_{\mu\nu} \rightarrow {B}_{\mu\nu}= b_{\mu\nu} +
 F_{\mu\nu}.}
Using the conditions on the background fields, it is easy to show that
\eqn\siete{\CL_k \, B_{\mu\nu} = 0.}
In order to perform the duality transformation, it is convenient to
rewrite \uno\ as a redundant gauge system where the isometry \dos\ is
gauged. We must introduce a Lagrange multiplier to ensure that the
auxiliary gauge field is flat. According to \siete, minimal coupling 
  is enough to construct the gauged action:  
\eqn\ocho{\eqalign{S_{\rm gauged} =& 
{1\over 4\pi} \int_{\Sigma} (g_{\mu\nu}\eta^{ab} +
 i\,{B}_{\mu\nu}\epsilon^{ab})D_a
x^{\mu}D_b x^{\nu}\cr & +{i\over 4\pi} \int_{\Sigma} 
{\tilde x}^{0}(\partial_a V_b - \partial_b V_a)
\epsilon^{ab}-{i\over 2\pi}\int_{\partial\Sigma}{\tilde x}^0 V,}}
where $D_a x^{\mu} = \partial_a x^{\mu} + k^{\mu}V_a$ and,
 in adapted coordinates the Killing vector reads
 $ k = {\partial\over \partial x^0}$.  
The one-form associated to the gauge field is represented as 
$V = V_a dx^a$.  
The r\^ole of the boundary term is to convey invariance under 
translations of the Lagrange multiplier: ${\tilde x}^0 \rightarrow
{\tilde x}^0 + C$.

Boundary conditions are restricted by several physical requirements.
The gauge parameter must have the same boundary conditions
as the world sheet fields (i.e. Neumann), in order for the isometry
to be realized on the boundary of the Riemann surface.
This has the obvious consequence that, if the gauge 
$V=0$ were needed, it would be neccessary to impose on 
the gauge fields the boundary condition $ n^{a}V_{a} \equiv V_n =0$
 (because this
component can never be eliminated with gauge transformations
obeying Neumann boundary conditions). It turns out, however, that
in order to show the equivalence of (2.7) with the original model
 (2.1), the behavior of $V|\,_{\partial\Sigma}$ is inmaterial.
The only way the action \ocho\ can now lead to the {\it unique} 
restriction $dV=0$ on the gauge field is to restrict the variations
of the Lagrange multiplier in such a way that $\delta {\tilde x}^0|\,_{\partial \Sigma}=0$.
In this way we are forced to impose Dirichlet boundary conditions
${\tilde x}^0 = C$ on the multiplier. Since the rest of 
the coordinates remain Neumann, a Dirichlet $(d-2)$-brane is obtained.
Besides, this ensures gauge invariance of \ocho.
\par
The last two terms in \ocho\ can be combined into
\eqn\ochoo{{i\over 2\pi}\int_{\Sigma}-d{\tilde x}^0 \wedge V.}
This means that the gauge field enters only algebraically 
in the action,
 and it can be replaced by its classical value
 (performing the gaussian integration only modifies 
the dilaton terms): 
\eqn\nueve{V_a^{c\ell}=-{1\over k^2}\left(
k^{\mu} g_{\mu\nu}\partial_a x^{\nu}
+i \epsilon_a\,^b \partial_b {\tilde x}^0 + i\, \epsilon_a\,^b k^{\mu}
B_{\mu\nu} \partial_b x^{\nu}\right).}
Particularizing now to adapted coordinates, $k = {\partial\over\partial x^0}$
and choosing the dual gauge $x^0 = 0$, we get
 the dual model, whose functional form is exactly like in \uno,
 but with the backgrounds ${\tilde G}_{\mu\nu}$, ${\tilde B}_{\mu\nu}$
 given in terms of the original ones 
through Buscher's formulas:
\eqn\b{\eqalign{{\tilde G}_{00} =& {\tilde g}_{00} = {1\over g_{00}}\cr
{\tilde G }_{0i} =& {B_{0i}\over g_{00}}  \cr 
{\tilde G}_{ij} =& g_{ij} -{g_{0i}g_{0j}-B_{0i}B_{0j}\over g_{00}}  \cr
{\tilde B}_{0i}=& {\tilde b}_{0i} = {g_{0i}\over g_{00}} \cr  
{\tilde B}_{ij}=& B_{ij}-{g_{0i}B_{0j}-B_{0i}g_{0j}\over g_{00}}.}}

In deriving these expressions, some care must be
exercised in choosing the appropriate variables in adapted
coordinates. In this frame, the isometry is represented by simple
translations $x^0 \rightarrow x^0 + \epsilon$. This means that the
various backgrounds must be independent of $x^0$, up to target space
gauge transformations, which in this model are defined by
\eqn\gau{ b\rightarrow b+ d\lambda \,\,\,\,\,\,\,\,\,\,\, A\rightarrow
A-\lambda ,}                                            
where $\lambda$ is an arbitrary one-form, in such a way that $B=b+dA $
is invariant. This gauge ambiguity is responsible for the
occurence of the non-trivial Lie derivatives in \tres. In order to 
consistently reach the gauge $x^0 =0$ in the closed string sector   
(world sheets without                                              
boundaries), the torsion Lie derivative   must be cancelled          
within the local patch of adapted coordinates. In fact, it
   is easily seen that                 
the gauge transformation $\lambda$, defined as                     
$$                                                                  
\CL_k \,\lambda = -\omega + d\varphi                                
$$                                                      
cancels                                                  
both the $\omega$ and $\varphi$ terms  in                          
 \tres. In this gauge, all fields are   locally independent of $x^0$.

 The behavior of the dilaton under T-duality is always a subtle
issue. In the present situation this is even more so, due to the fact
that the metric is not a massless background of the open string, and
one has to consider closed string corrections, thus driving the sigma
model away from the conformally invariant point, in order to get a
consistent Fischler-Susskind mechanism (cf. \ref\clny{C.G. Callan, C.
Lovelace, C.R. Nappi and S. A. Yost, \npb{288} {1987} {525};
\npb{293} {1987} {83}; \npb{308} {1988} {221 \semi} 
J. Polchinski and Y. Cai, \npb{296} {1988} {91.}} 
    and references
therein). There is, however, a neccesary condition for the equivalence
of the two theories, and this is that the effective action must remain
invariant. It turns out that this condition is sufficient to determine
the dual dilaton to the value
\eqn\dil{{\tilde \phi} = \phi - {1\over 2} \,{\rm log}\, k^2 . }

The invariance under translations of the dual model can sometimes be
put to work in our benefit. Let us consider, for simplicity, the
Wilson line $A_0 = {\rm diag} (\theta_1, ..., \theta_N)$, when only
the coordinate $x^0$ is compactified in a circle of length $2\pi R$ in
an otherwise flat background. The Wilson line itself, in the sector of
winding number $n$ is given by
$$
\sum_{a=1}^N e^{2\pi nR\theta_a i   }
$$
This term is reproduced in the dual model
 by simply  taking into account
 the ``total derivative term" coming from the gaussian
integration of the auxiliary gauge field, namely
$$
\int_{\Sigma} dx^0 \wedge d{\tilde x}^0
,$$
which by Stokes' theorem can be written as $-\oint {\tilde
x}^0 \, dx^0 = - C\,2\pi nR$, where $C$ is the constant value of the
multiplier on the boundary. This value can depend on the (implicit)
Chan-Paton indices of the world sheet fields, and we recover with our
techniques the result of      \Topen\po\P\
 that the Wilson line considered
induces in the dual model a series of D-branes with fixed positions
determined by the $\theta_a$ parameters. A more detailed discussion,
generalized to arbitrary genus world sheets can be found in appendix
A.

An interesting observation is that the collective motion of the
D-brane is already encoded in Buscher's formulas. To see this, notice
that the dual backgrounds in \b\ differ from the standard duals
without gauge fields by the terms:
\eqn\induc{
\eqalign{ {\tilde G}_{0i} = & \tg_{0i} -\tg_{00} \, \partial_i A_0 \cr
\tG_{ij} =& \tg_{ij} + \tg_{00} \, \partial_i A_0 \partial_j A_0 -
\tg_{0i} \, \partial_j A_0 - \tg_{0j} \, \partial_i A_0 \cr
\tB_{ij} =& \tb_{ij} + F_{ij} + \tb_{0i} \, \partial_j A_0 - \tb_{0j}
\, \partial_i A_0 . }}
  Using these formulas it is easily checked that the dual
 sigma model reduces to the standard one  in terms of the backgrounds
$\tg_{\mu\nu}$ and $\tb_{\mu\nu}$, provided we make the replacement
\eqn\shift
{\tx^0 \rightarrow \tx^0 + A_0 (x^i).}
Thus, the gauge field component $A_0$ acquires the dual interpretation
of the transverse position of the D-brane, as a function of $x^i$,  
which become longitudinal world volume coordinates\foot{It is amusing
to notice that, if we choose to cancel only the $\omega $ dependence
in \tres, then the previous formulas still hold if we substitute $A_0$
 by $A_0 - \varphi$. In addition, we have $\pt_0 A_j = \pt_j \varphi$.
If we were so bold as to interpret $x^0$ as a timelike coordinate, we
could regard the $A_j$ configuration as a dyonic field.}. 
 The same result
follows from careful consideration of the boundary conditions.

\subsec{Boundary Conditions}
It is well known \clny\  
  that in 
the presence of nontrivial backgrounds the Neumann boundary
 condition, which physically is equivalent to the absence 
of momentum flow through the edges of the string,
 is modified (because the physical momentum is modified accordingly), to
\eqn\diez{g_{\mu\nu} \partial_n x^{\nu} + i B_{\mu\nu}
 \partial_t x^{\nu} = 0,}
where the vector $n$ and $t$ are the normalized outer
 normal and tangent vector
to the world sheet boundary   and,
 correspondingly, $\partial_n \equiv n^a \partial_a$,
 and $\partial_t \equiv t^b \partial_b$.  

It is not immediatly clear what are the correct generalizations of \diez\
 for the gauged action \ocho. We claim that the good choice is:
\eqn\once{g_{\mu\nu}D_n x^{\nu} +i B_{\mu\nu}D_t x^{\nu}=0.}
First of all, if we use the fact that the gauge connection is flat,
$V_a = \partial_a \alpha$, and perform an isometric transformation
$x^{\mu}\rightarrow x^{\mu}+ \alpha k^{\mu}$, then 
 \once\ easily yields \diez.
On the other hand, \once\ reduces, in adapted coordinates
 and in the $x^0 = 0 $ gauge, to
\eqn\trece{\eqalign{& g_{00}V_n + g_{0j}
 \partial_n x^{j} + i\, B_{0j} \partial_t x^{j}=0 \cr 
& g_{i0}V_n +g_{ij}\partial_n x^{j} +
 i\, B_{i 0} V_t + i B_{ij} \partial_t x^{j} = 0.}}
If now the classical gauge fields $V_a^{c\ell}$ are plugged
 in \trece, and use is made of Buscher's formulas, the following
 dual boundary conditions are obtained:
\eqn\quince{\eqalign{&\partial_t {\tilde x}^0 = 0 \cr
&{\tilde G}_{i \mu} \partial_n \tx^{\mu} +
 i\, {\tilde B}_{i \mu}\partial_t \tx^{\mu} = 0.\cr}}
where we denote $\tx^{\mu} = (\tx^0, x^j)$.
It is also possible to verify involution ($T^2
= 1$).
The dual model is automatically written in adapted coordinates 
to the dual Killing $ {\tilde k}= {\partial \over \partial 
{\tilde x}\,^0}$.
A further T-duality transformation can then be performed with respect to it.
Substituting again the values of the
  dual classical gauge fields,
${\tilde V}_a^{c\ell }$, in the dual covariant boundary conditions, the correct ones
 for the original model \diez\ are recovered.  

The mixed boundary conditions \quince\ correspond to a Dirichlet
$(d-2)$-brane in the dual background. To check this,  notice that
 for a
D-brane of general shape given by the equations   
$$
\tx^{\mu} = y^{\mu} (x^j)
,$$
we can define induced metric and antisymmetric tensor backgrounds in
the usual fashion
\eqn\indu
{(\tg_{ij})_{\rm induced} = \partial_i y^{\mu}\,
 \tg_{\mu\nu}\, \partial_j
y^{\nu},} 
and similarly for $\tb_{\rm induced}$. As expected, the previously
defined $\tG_{ij}$ and $\tB_{ij}$ in \induc\ are precisely the induced
world volume backgrounds under the identification $y^0 = -A_0$, $y^j =
x^j$. Furthermore, shifting the dual coordinate $\tx^0 \rightarrow
\tx^0 + A_0$ we obtain the dual boundary conditions in the form of
the second reference in  \Topen: 
\eqn\ot{
\eqalign{&  \tx^0 = y^0 (x^j) \cr 
&\partial_i y^{\mu} \,\tg_{\mu\nu} \partial_n \,
 \tx^{\nu} + i\, \tB_{i\nu}
\,\partial_t \, \tx^{\nu}=0. }}

In the case of unoriented strings, the change from Neumann to
Dirichlet conditions is supplemented by an orbifold projection in the
space-time which reverses the orientation of the world sheet (the 
orientifold). This result follows easily  from the
T-duality mapping at the level of the conformal field theory (see
\Topen ):
$\partial_z x(z) \rightarrow \partial_z x(z)$, $\partial_{\bar z}
x({\bar z}) \rightarrow - \partial_{\bar z} x({\bar z})$. In order to
study this question in a curved background, let us consider the 
 lowest order unoriented  topology,
 namely the crosscap  $P_2(R)$. It is enough for our purposes to make
 in the boundary of the unit disc the identification of opposite points:
$x^{\mu} (\sigma) = x^{\mu} (\sigma+\pi)$. (Here $\sigma \in (0,2\pi)$
just parametrizes the boundary.)
Dropping the zero mode,
this yields the conditions  
 \clny:   
\eqn\acero{\eqalign{ \partial_n \, x^{\mu}(\sigma)=& -\partial_n \,
x^{\mu}(\sigma + \pi)  \cr 
\partial_t\,  x^{\mu}(\sigma) =&\,  \partial_t 
\, x^{\mu}(\sigma + \pi).\cr}}

When gauging the isometry, there 
is a covariant generalization of these conditions, namely
\eqn\atres{\eqalign{D_n x^{\mu}(\sigma) =& - D_n x^{\mu}(\sigma + \pi)
\cr
D_t x^{\mu}(\sigma) =&\, D_t x^{\mu}(\sigma + \pi).}}
Using the value of $V_a^{c\ell}$ obtained above, we easily find, 
after fixing the $x^0 =0$ gauge: 
\eqn\ds{
D_n x^0 = - {1\over k^2}\partial_n x^i k_i +i\, \partial_t {\tilde x}^0
.}
The antisymmetric tensor and abelian gauge field backgrounds are
projected out from the physical spectrum of unoriented strings in the
weak field limit, and so we only consider a nontrivial metric
background. Then \atres\ and \ds\ 
 yield the dual boundary conditions in the form:
$$\left(i\,\partial_t \tx^0 -{1\over k^2} k_j \partial_n x^j\right)
(\sigma+\pi)=
 -\left( i\,
\partial_t{\tilde x}^0 - {1\over k^2} k_j \partial_n x^j\right)
(\sigma).$$
The terms containing the Killing vector cancel away
 owing to the boundary
 conditions of the original model;
 the rest reduces to the orientifold condition on $\tx^0$:  
\eqn\atress{\eqalign{\partial_t  {\tilde x}^0 (\sigma) =&
-\partial_t  {\tilde x}^0 (\sigma + \pi) \cr   
\partial_n \tx^0 (\sigma) =&\, \partial_n \tx^0 (\sigma + \pi) .}} 

The first equation implies $\tx^0 (\sigma) + \tx^0 (\sigma + \pi) =
{\rm constant}$, so that the crosscap is embedded as a twisted state
of the orbifold. 
This is one of the main results of this paper; it implies that 
the orbifold character of the dual target space is
 a generic phenomenon, and not
 a curious peculiarity of toroidal backgrounds.
It is curious to remark that the dual manifold always enjoys
parity ${\tilde x}^0 \rightarrow -{\tilde x}^0$ as an
isometry, because ${\tilde g}_{0i} =0$.

The rest of the coordinates still satisfy standard crosscap
conditions. An important consequence of 
 \atress\ is that               
at least two  points of the boundary 
are mapped to the orientifold fixed points
in the target, which means that local contributions of 
non orientable world sheets 
are concentrated at  the orientifold location;  
 in the bulk of space-time the
dual theory is orientable along the direction $\tx^0$ \Topen.  
This is compatible with the appearance of a non vanishing dual
antisymmetric tensor $\tb_{0i} = g_{0i} /g_{00}$ as long as the
original background has a ``boost" component. The effects of this
background field are supressed only for world sheets mapped to the
fixed point. Another observation is that, in the absence
of a $U(1)$ gauge field, there is no collective coordinate for the
orientifold, which becomes a rigid object. Indeed, according to the
formulas \induc\ and \indu, the induced backgrounds are exactly the
same as the vacuum dual backgrounds.

\newsec{Superstrings  }

The extension of the previous results to the supersymmetric case is
for the most part straightforward. In particular, the dual background
formulas \b\ still apply. There are, however, some subtleties
associated with the choice of fermion boundary conditions in the
presence of non trivial torsion or abelian gauge fields.  
We find that, generically, T-duality induces complicated boundary
conditions mixing fermions and bosons.

We shall examine these problems  for the general $(1,1)$
  supersymmetric sigma model in the lowest order world sheets (the
disk and the projective plane). The action is   
\eqn\ctres{S =- {1\over 2\pi} \int d^2 z\, d^2 \theta\,\,
 (g_{\mu\nu} + B_{\mu\nu})\,D_{+}X^{\mu}D_{-}X^{\nu}.}
where  the superfields are  $X^{\mu}(\theta_{+},
\theta_{-},z,\bar z) = x^{\mu} + \theta_{+}\psi_{+}^{\mu} +
 \theta_{-}\psi_{-}^{\mu} + \theta_{+}\theta_{-}F^{\mu}$,
 the superderivatives 
being $D_{+} = \partial_{+} +
 \theta_{+}\partial_{z}$ and $D_{-} =
 \partial_{-} + \theta_{-}\partial_{\bar z}$.  
When expanded in components, \ctres\ contains the standard interaction
terms of the general $(1,1)$ sigma model with torsion, and also a
fermionic boundary action of the form
\eqn\bou{ S_{\rm line} = {i\over 4\pi} \oint B_{\mu\nu} \left(
\psi_+^{\mu} \psi_+^{\nu} + \psi_-^{\mu} \psi_-^{\nu}\right). } 
The gauge component (recall that $B= b+dA$) is the well known
fermionic part of the supersymmetric (abelian) Wilson line.

The model \ctres, with $d$-dimesional target 
space would appear in the discussion of type II strings
with Dirichlet $(d-1)$-branes 
in general (longitudinal) background fields. Also, if we
drop the torsion and abelian gauge fields $B=0$ we  may consider a
type I string in curved space\foot{In type I superstrings, the only
consistent backgrounds are Wilson lines of $SO(32)$, which need a
different treatment (see appendix A for a discussion of some simple
cases). In particular, the dual contains an orientifold and 16
D-branes. It is important, however to consider general backgrounds with
torsion because even when it is zero in the initial theory, it is
generically induced in the dual one after a T-duality
transformation.}.
In order to keep the discussion as simple as possible,
 we start with generalized
boundary conditions of Neumann type in all coordinates in \ctres, and
define the mixed boundary conditions at the dual $(d-2)$-brane as
those induced by T-duality. In the bosonic sector, Neumann boundary
conditions are defined enforcing the complete boundary equations of
motion. However, in the presence of fermionic boundary actions such as
\bou, this requirement might be too strong, since it would produce
trivial quantum dynamics of the boundary fermions.     

This is easily illustrated in     the simple situation where
$g_{\mu\nu} = \delta_{\mu\nu}$  and $B_{\mu\nu} =b_{\mu\nu}
$ is constant. Using matrix notation, 
the  full action  splits  into a {\it surface} and a {\it line} term
\eqn\bulk{\CL_{\rm surface} = -\pt_z x^t (1+b)\, \pt_{\zb}
 x + \psi_+^t \pt_{\zb} \psi_+ +
\psi_-^t \pt_z \psi_- }
\eqn\borde{\CL_{\rm line} = {1\over 2} \pt_{\zb} ( \psi_+^t b\, \psi_+) - {1\over 2}
\pt_z (
\psi_-^t b\, \psi_- ).}
It is important to notice that $\CL_{\rm surface}$ enjoys invariance under
chiral rotations of the form $\psi_- \rightarrow \CO \psi_-$
provided  $\CO^t \CO = 1$ is orthogonal. 
If we want to promote this
to a symmetry of the full action (including the {\it line} term),
then we have to demand in addition that
 $[\CO, b]=0 $. 

The variation of the full action induces the following boundary terms
\eqn\neu{
\oint \delta x^t \left( (1-b)\,\pt_z x - (1+b)\, \pt_{\zb} x\right) +\delta
\psi_-^t (1+b)\, \psi_- - \delta \psi_+^t (1-b)\, \psi_+
,}
which lead to the standard \clny\  conditions on the bosonic sector,
\eqn\usual
{\pt_z x = {1+b \over 1-b}\, \pt_{\zb} x .}
However, neither the usual NS-R conditions $\psi_+ = \eta\, \psi_-$,
$\eta=\pm 1$, nor the ``rotated" ones
\eqn\rota{ \psi_+ = \eta \, {1+b\over 1-b}\, \psi_- ,}
solve \neu. Notice that \rota\ is equivalent to the normal NS-R
conditions by means of the above mentioned chiral rotation symmetry,
with $\CO =  (1+b)/(1-b)$.
The boundary conditions \usual\ and \rota\ are the ones considered in
\clny, using the operator formalism.

An explicit solution of \neu\ is given  by  the ansatz
\eqn\fermans
{\psi_+ = \eta\, {1+b \over 1-b}\, \CJ \psi_-}
 provided  $\CJ^t \, (1-b) \,\CJ
 = (1+b)$. The matrix $\CJ$  can be chosen
satisfying $\CJ^t = \CJ$,
 $\CJ^2 = 1$, and $\{\CJ, b\} =0$. 
If  $b$ is block-diagonalized in the form $b = {\rm diag}_j
 (ib_j \sigma_2)$,
then  $\CJ ={\rm diag}_j (\sigma_3)$ (notice the resemblance with 
 an (almost) 
 complex structure
in the target).  
Given that $(1-b)/ 
(1+b)$ is orthogonal and commutes with $b$, the boundary conditions
\fermans\ are a chiral rotation of 
 $\psi_+ = \eta \,
 \CJ\,\psi_-$, which explicitly break the $O(d)$ Lorentz symmetry, even
in the limit of vanishing electromagnetic background.

The important point is that with the ``fully classical" boundary
 conditions
\fermans\ 
$$
S_{\rm line} \sim \oint (\psi_+^t b\, \psi_+ + \psi_-^t b\, \psi_-) =0
,$$
as corresponds to an on-shell first order system. 
Therefore,
 in order to have some quantum  fermionic dynamics on the boundary,
we may set to zero  only those total derivatives  
 coming from the variation of
$\CL_{\rm surface}$, letting $\CL_{\rm line}$ freely fluctuate.
In this way we are led to the standard NS-R conditions in the
fermionic sector, which we can rotate by a convenient orthogonal
matrix, thus recovering  \rota. The particular combination of
 the $(1,1)$ world sheet supersymmetry
which is preserved depends on such rotations. The variation of the
complete action 
 under transformations generated by the currents
$ J_+ = -\psi_+^t \, \pt_z x(z)$, and $J_- = -\psi_-^t \, \pt_{\zb}
x(\zb)$ has the form
$$
\delta_{\rm ss} S \sim \oint \left(\epsilon_+ \psi_+^t - \epsilon_-
\psi_-^t \right) (1-b) \, \pt_z x
,$$
where we have used the bosonic boundary conditions \usual. We see that
the NS-R conditions $\psi_+ = \eta\, \psi_-$ ensure invariance under
the supersymmetry generated by $J_+ + \eta\, J_-$, whereas the rotated
 ones  \rota\
preserve instead the conveniently rotated current
$$
J_{\rm rot} = -\psi_+^t \, \pt_z x -\eta\, \psi_-^t \,
{1-b\over 1+b}\, \pt_{\zb}
x
.$$
These currents define the induced supersymmetry on the boundary. 
Note that the surviving boundary supersymmetry is not quite the same
as the combination that maps the fermionic and bosonic conditions
into each other. For example, $J_+ + \eta\, J_-$ maps \rota\ into
\usual.

For curved  backgrounds, a very general class of boundary
conditions with consistent T-duals is given by:       
\eqn\conds{
\eqalign{ (g-B)_{\mu\nu} \pt_z x^{\nu} - (g+B)_{\mu\nu} &  \pt_{\zb}
x^{\nu} + W_{\mu} (\psi_-) =0   \cr
\psi_+^{\mu} =& R^{\mu} (\psi_-) .}} 
The functionals $W$ and $R$ are regarded as power series expansions in
the right moving fermions $\psi_-^{\mu}$, with coefficient  functions
satisfying
the isometry constraint (independent of $x^0$). There are two
particular solutions of special interest, corresponding to the two
classes of boundary conditions considered previously. 
 With the choice $W_{\mu} =0$
and $R^{\mu} (\psi_-) = \eta \, \psi_-^{\mu}$, we 
recover the standard Neumann NS-R conditions
 which we would derive neglecting the boundary variations of \bou. On
the other hand, a set of ``fully classical" boundary conditions
(including \bou), can be obtained setting
 $R^{\mu} (\psi_-)  = \eta \, \CJ^{\mu}_{\nu} \psi_-^{\nu}$,
and $W_{\mu} (\psi_-) = \Omega_{\mu\nu\rho} 
\, \psi_-^{\nu} \psi_-^{\rho}
$, with the constraints 
$$
\CJ^{\mu}_{\alpha} (g-B)_{\mu\nu} \, \CJ^{\nu}_{\beta} =
(g+B)_{\alpha\beta},$$
$$
\Omega_{\mu\nu\rho} = B_{\mu [\nu;\rho]} + B_{\alpha\mu ; \beta}
\,\CJ^{\beta}_{[\nu} \, \CJ^{\alpha}_{\rho]} + B_{\alpha\beta} \,
\CJ^{\alpha}_{[\nu} \, \CJ^{\beta}_{\rho] ;\mu} . $$ 
It is easily checked that these twisted conditions still satisfy
$S_{\rm line} =0$.

Now the standard algorithm for T-duality goes through in the
superfield formalism. 
The action is invariant under the super-isometry transformation
$ \delta_{\epsilon} X^{\mu} = \epsilon \, k^{\mu}$,   
provided the background fields satisfy 
the same conditions as in the bosonic case.
The auxiliary gauging  is achieved again by minimal 
coupling to the gauge superfields $V_{+}$ and $V_{-}$,  
whose transformation laws  under the gauged isometry are 
 $\delta_{\epsilon} V_{\pm} = -D_{\pm}\epsilon$,  
where now $\epsilon$ is a $(1,1)$  bosonic superfield.
The other components of the supergauge multiplet are 
written in terms of $V_{\pm}$ as $V_{z} =
 D_{+}V_{+}$ and $V_{\bar z} = D_{-}V_{-}$.
The supercovariant derivatives are $\nabla_{\pm}X^{\mu} =
 D_{\pm}X^{\mu} + V_{\pm}k^{\mu}$ and $\nabla_{z(\bar z)}X^{\mu} = 
\partial_{z(\bar z)}X^{\mu} + V_{z(\bar z)}k^{\mu}$.  
The gauged action with Lagrange supermultiplier term is
\eqn\cocho{S_{\rm gauged}  = -{1\over 2\pi} \int d^2 z \, d^2 \theta\,
\left( (g_{\mu\nu} + B_{\mu\nu})\nabla_{+}X^{\mu}\nabla_{-}X^{\nu}
- D_+ \tX^0 \,V_{-} - D_{-} \tX^0 \, V_{+}\right).}
Integrating out the supermultiplier enforces the super-flatness
condition $D_+ V_- + D_- V_+ =0$, ensuring the equivalence of
\cocho\ and \ctres. If the gauge superfields are integrated instead,
the dual model follows after gauge fixing $X^0 = 0$.

The gauged boundary conditions, equivalent to \conds,  can be written as
\eqn\cnueve{\eqalign{g_{\mu\nu}(\nabla_{z} - 
\nabla_{\bar z})X^{\nu} - B_{\mu\nu}(\nabla_{z} + 
& \nabla_{\bar z})X^{\nu} + W_{\mu} (\nabla_- X)  
\vert_{\theta = 0} =0 \cr  
\nabla_{+}X^{\mu} =& R^{\mu}
 (\nabla_{-}X) \vert_{\theta = 0}.}} 
The transformation to the  dual variables proceeds as usual by 
 fixing the gauge $X^0 =0$ in adapted coordinates. The final result is
simply obtained by the substitutions $\pt_z x^0 \rightarrow D_+
V_+^{c\ell}$, $\pt_{\zb} x^0 \rightarrow D_- V_-^{c\ell}$, and
$\psi_{\pm}^0 \rightarrow V_{\pm}^{c\ell}$ in eq. \conds, with the
following classical gauge supermultiplet:  
$$ 
D_{\pm} V^{c\ell}_{\pm} \vert_{\theta=0}  =\pm\pt_i \, {\tilde k}^{\mp}_j \,
\psi_{\pm}^i \psi_{\pm}^j \pm \pt_i \,{\tilde k}^2 \, \psi_{\pm}^i
{\tilde \psi}_{\pm}^0 \pm 
{\tilde k}^{\mp}_i \pt_{z(\zb)} x^i \pm {\tilde k}^2
\pt_{z(\zb)} \tx^0
$$
$$
V^{c\ell}_{\pm} \vert_{\theta=0} = \pm{\tilde k}^{\mp}_i \, \psi_{\pm}^i \pm
{\tilde k}^2 {\tilde \psi}_{\pm}^0 ,
$$
where ${\tilde k}^{\pm}_{\mu} \equiv {\tilde k}^{\nu} (g\pm
B)_{\nu\mu}$. One finds a complicated set of mixed Dirichlet-Neumann
conditions, with the same purely bosonic terms as in (2.18),  and 
extra  boson-fermion interactions.  
 A rather non
trivial check, which is   met by  these duality mappings,  
 is provided by the involution
property: $T^2 =1$.

Explicit (somewhat complicated) expressions 
 for the dual functionals ${\tilde
R}^{\mu}$ and ${\tilde W}_{\mu}$ can be read off from the previous
manipulations in all generality.  Here we will just quote the 
explicit form of the simplest (and perhaps more interesting) boundary
conditions: the  
Neumann NS-R case (that is $W=0$ and $R=\eta$
in \conds),
\eqn\conce{\eqalign{\tilde{\psi}_{+}^i =& \eta\,  \tilde{\psi}_{-}^i
\cr 
\tilde{\psi}_{+}^0 +\eta\,  \tilde{\psi}_{-}^0 =& -2\,\eta\, k^*_j
\, { \tilde \psi}^j  \cr  
(\pt_z + \pt_{\zb})\tx^0 =& -2\,\pt_i \, k^*_j
\,\tpsi_-^i \tpsi_-^j \cr  
 \tg_{i\mu}
 (\pt_z - \pt_{\zb})\tx^{\mu} - {\tilde B}_{i\mu} (\pt_z +
\pt_{\zb})\tx^{\mu}   = 2\, k_i^* \,\pt_j \,
{\tilde k}^2 \,
 &\tpsi_-^j \tpsi_-^0 - 2\,({\tilde k}_i^+ \,\pt_j \,k_l^* - k_i^* \,
\pt_j
\,{\tilde k}_l^+ )\, \tpsi_-^j \tpsi_-^l }}
where
we denote
  $k_i^* \equiv {\tilde k}_i /{\tilde k}^2 = B_{0i}$, and $\tpsi^i
= \psi^i$.  
Notice that the non trivial
fermionic 
terms are all proportional to the ``electric field" $B_{0i}$, and 
therefore correspond to a certain D-brane shape and state of motion
       in the dual language.
In particular, they are absent for the case of type I/8-brane (with
orientifold)  duality.
Using \siete\ and the Bianchi identities in 
 the pure electromagnetic case $B=dA$,
it is easy to check that 
the fermionic terms drop from the Dirichlet condition of the
multiplier $\tx^0$, in the third equation of \conce.    
The
relative minus sign in the induced spin structure of the fermions
$\tpsi_{\pm}^0$ accounts for the reversal of space-time chirality: if
the closed string sector of the original model is type A, then the
T-dual is type B. 

 In the case of the type I string we have to consider the projective
plane to leading order.
In fact, the absence of torsion in the original background makes the
discussion much simpler. 
Let us denote by $ P = \sigma $ and $ P' = \sigma + \pi$
the two identified points on the boundary of the disk.
The standard  crosscap bosonic  conditions \acero, together with
$\psi_+^{\mu} (P') = \eta \, \psi_-^{\mu} (P)$, are equivalent to the
covariant ones     
\eqn\cov{\eqalign{\nabla_{+} X^{\mu}(P') =& \eta \,
   \nabla_{-} X^{\mu}(P)|\,_{\theta =0} \cr  
(\nabla_z - \nabla_{\bar z})X^{\mu}(P') =
& (\nabla_z -\nabla_{\bar z}) X^{\mu}(P)|\,_{\theta =0} \cr  
(\nabla_z+\nabla_{\bar z})X^{\mu}(P') =& - (\nabla_z + \nabla_{\bar z})
X^{\mu}(P)|\,_{\theta = 0}.}}
In the gauge $X^0 = 0$ they reduce to
\eqn\hh{\eqalign{
{\tilde \psi}_{+}^0 (P') =& -\eta \,  {\tilde \psi}_{-}^0(P) \cr  
(\partial_z - \partial_{\bar z}){\tilde x}^0 (P') = 
& - (\partial_z - \partial_{\bar z}){\tilde x}^0 (P) \cr  
(\partial_z + \partial_{\bar z}) {\tilde x}^0 (P') =
& (\partial_z + \partial_{\bar z}) {\tilde x}^0(P),}}
getting in this way the orientifold conditions for the boson as well as the corresponding
change of sector for the fermion, the remaining equations being unaltered.

It would be interesting to study the interplay between \conce\ and
space-time supersymmetry in the context of $(2,2)$ K{\"a}hler sigma
models. In particular, for isometries acting without
fixed points, we expect no subtleties in the T-duality mapping of
target space supersymmetry charges \aagb. The total or partial  
 supersymmetry breaking
induced by the ``boost" terms in \conce\ (proportional to $k^*_j$),
reflects again the presence of non 
 vanishing ``electric fields" $B_{0i}$ in the original
model.    
\bigskip

\newsec{T-duality of the effective world volume actions}

In the closed string sector, the T-duality transformations \b,
together with \dil,  leave
 the low energy effective action invariant (in the string
frame). Accordingly, in  the presence of open string
backgrounds, T-duality should produce directly the Dirac-Born-Infeld
action appropriate for the corresponding D-brane. 
We may illustrate this fact for the simplest case. We start from the
ten dimensional open string effective action, to leading order in a
derivative expansion:  
\eqn\teneff{
S_{\rm eff} = \int {d^{10}x \over \alpha'^5}\, e^{-\phi}\,\sqrt{ 
 {\rm det}
(g+b+F)_{\mu\nu}}
.}
For the unoriented type-I string, $b=0$ and $F$ lies in $SO(32)$, so
that a trace over group indices is understood.

We 
will consider factorized Chan Paton factors such that $A_0$ is
diagonal and $A_i$ is independent of $x^0$ and assumes a block
factorized form according to the eigenspaces of $A_0$.
 Then $ F_{0j} = \partial_0 A_j - \partial_j A_0 =
 -\partial_j  A_0 = \partial_j \, y $.  
The basic identity we need is the following, which holds for an
arbitrary matrix $M_{\mu\nu}$: 
$$
{\rm det}(M_{\mu\nu}) = M_{00} \, {\rm det}\left( M_{ij} -
{M_{i0}M_{0j} \over M_{00}} \right)
.$$
 Applying the formula to the determinant
above we find
$$
{\rm det}(g+b+F)_{\mu\nu} = g_{00} {\rm det}
[({\tilde g}+{\tilde b} )_{ij} +F_{ij} +{\tilde g}_{00} \partial_i y 
\partial_j y +\partial_i y ({\tilde g} +{\tilde b})_{0j} +\partial_j
y 
({\tilde g} - {\tilde b})_{0i}] 
$$
where ${\tilde g}_{\mu\nu}$ and ${\tilde b}_{\mu\nu}$ are the standard
vacuum dual fields. Using \induc\ and \indu, we easily obtain the
world volume action for the dual 8-brane:
\eqn\ochob{
\int {d^{10} x \over \alpha'^5} \, e^{-\phi} \, \sqrt{ {\rm
det}(g+b+F)_{\mu\nu}} = 2\pi \int {d^{9} x \over  \alpha'^{9/2}} \,
e^{-{\tilde \phi}} \, \sqrt{ {\rm det}(({\tilde g} + {\tilde b})_{\rm
induced} + F)_{ij}}
,}
where we have used the relations $\int dx^0 = 2\pi \sqrt{\alpha'}$ and
${\tilde \phi} = \phi - {\rm log}\,\sqrt{g_{00}}$. 
Notice that the dilaton transformation is essential to cancel the
Kaluza-Klein factor of the compact volume, so that we obtain the
correct dimensionality for the world-volume action. We see that the
results of \Topen\ follow
 inmediately from those of \clny\  using
T-duality.

In the context of type I and II strings, we have also R-R $p$-forms
$A_p$, whose transformation law under duality is not obvious from the
sigma model construction, because R-R backgrounds involve vertex
operators with non-trivial spin field and super-ghost dependence. It
is possible to guess their transformation laws  requiring
T-duality of effective low energy effective action 
 \ref\tomas{E. Bergshoeff, C. Hull and
T. Ort{\'{\i}}n, \npb {451} {1995} {547.}}.
Here we just check the T-duality covariance of the generalized D-brane
Chern-Simons actions in the simplified case where $g_{0i} = B_{0i}
=0$.
 The universal coupling of the R-R forms to the $p$-brane is
 \clny\ \ref\miao{M. Li, hep-th/9510161\semi 
C.G. Callan and I. Klebanov, hep-th/9511173.} 
\ref\dou{M. Douglas, hep-th/9512077\semi
C. Schmidhuber, hep-th/9601003.}  
\eqn\sour{
S_{\rm source}^{RR} 
  = (2\pi \sqrt{\alpha'})^{3-p} \int_{\Sigma_{p+1}}
\sum_{q=1}^{p+1} A_{q}\, {\rm Str}\, e^{b+F + dy}
,}
up to corrections in derivatives of the NS-NS  field strength $B=b+F$. 
``Str" is an antisymmetrized trace over
gamma matrices: we expand the exponent with the structure ${\rm exp}
\gamma^{\mu} \gamma^{\nu} B_{\mu\nu}$ and antisymmetrize all the
traces over gammas. The leading term is the usual
$\int_{\rm world-volume} A_{p+1}$, in   units   
$A_{p+1} \sim {\rm length}^{-4}$. 
The sum over $q$ in \sour\  runs over odd numbers for type IIA theory and
over even numbers for type IIB and type I (in fact, for type I there
are only one, five and nine branes\foot{Several consistency
requirements, most notably modular invariance and anomaly
cancellation, restrict the general structure of D-manifolds. See, for
example   
 \ref\gipo{E. Gimon and J.
Polchinski, hep-th/9601038.}.}).          
Since T-duality interchanges A and B types, the T-duality
transformation must be, to leading order:
$$
A_{\mu_1,\cdots,\mu_p,0} = {1\over \sqrt{g_{00}}}
 {\tilde A}_{\mu_1,\cdots ,
\mu_p}
.$$
So, a $p+1$ even form in IIB or type I goes into a $p$ odd form in
type IIA or type $I'$. With this form of the duality transformation
we ensure T-invariance of the kinetic term,  
$$
{1\over (p+2)!} 
\int d^{10} x \sqrt{g} |F_{p+2}|^2 ={1\over (p+1)!}
 \int d^{10} x \sqrt{\tilde g}
|{\tilde F}_{p+1}|^2
,$$ 
and the correct mapping  of the source terms. 
   $$
\int_{\Sigma_{p+1}} A_{p+1}= 
2\pi\sqrt{\alpha'} \int_{\Sigma_{p}} {\tilde A}_p
.$$

 It is important to point out that the duality formulas in this
section, particularly \ochob, do not take into account the back
reaction on the background fields caused by the presence of the
D-brane. In that case, an inhomogeneus profile in the $x^0$ direction
is generated for the
various fields, including the dilaton. Such non-trivial dependence on
$x^0$ spoils the isometry property, and the standard T-duality
formulas do not apply. These back reaction effects are important in
order to resolve some paradoxes in the type I/Heterotic string
duality \polwi. Large sources for the dilaton appear as large
uncancelled tadpoles in string perturbation theory. Such tadpoles are
homogeneus throughout space-time in the type I theory, and
inhomogeneus in the T-dual language, where there is an orientifold in
place. It would be interesting to understand the T-duality
transformation of the corrected backgrounds, even if they do not enjoy
an isometry symmetry. In other words, we would like to understand
T-duality in the presence of a non-trivial Fischler-Susskind
mechanism. A simple observation to make is that the isometry property
was used in \ochob\ only to represent the right hand side as a
nine-dimensional integral. If the $x^0$ integral is not simply
factorized in the compact volume, eq. \ochob\ still holds in the sense
of an ``average" over the compact dimension.    
       
\newsec{Conclusions}
Generic  T-duality transformations of  curved D-manifolds  with 
commuting isometries have been studied, and  Buscher's formulas 
have been shown to follow, including the collective motions of the
D-branes in the dual target space. We have also considered in detail 
the
T-duality between Neumann and Dirichlet boundary conditions in general
sigma models of bosonic or supersymmetric $(1,1)$ type.             
Orientifolds, in particular, have been obtained as  generic 
phenomena,
not tied up to specific aspects of toroidal compactifications.
\par
We have defered to the appendices a number of issues  of formal
nature. These include the consideration of higher genus world sheets,
albeit with the simplest backgrounds only (we can see no obstruction
in principle, however to repeat the analysis in the general case).  
Some aspects of very simple Wilson lines have also been studied,
although the
T-duality transformation of the most general non-abelian 
Wilson line remains as an interesting open problem.  
\par
 An appendix on T-duality in general, in the case in which
the orbits of the isometry group are non-compact,  has been included
 for pedagogical purposes, and some 
specific examples have been worked out in detail, 
although the results here are not new (cf. \ref\aagbl{E. Alvarez, L. 
      Alvarez-Gaum\'e, J.L.F. Barb\'on and Y. Lozano,
\npb {415}{1994}{71.}}).
Finally, we have outlined the canonical formalism for T-duality with
boundaries in the flat background. 

Much work remains to be done in the analysis of supersymmetric sigma 
models, specially those with extended supersymmetry, and the
interplay with space-time supersymmetry. 
 It is also important to  study  the effects of generic non-abelian 
gauge backgrounds, as well as space-time dependent dilaton sources,
whose consideration seems to be neccessary to really understand
the appearance of the strong couplig regime preventing  
a disproof of Heterotic/Type I string/string duality in \polwi.

\par
{\bf Acknowledgements}
We have benefited from discussions with L. Alvarez-Gaum\'e, C. Bachas,
C. Callan, C. G\'omez, Y. Lozano and  P. Townsend.
The work of E. A. is supported by the grant AEN/93/0673.
J.L.F.B. is supported by NSF PHY90-21984 grant. 
J.B. enjoys a CAM fellowship.

\appendix{A}{Global issues}

In this appendix  we give a path integral derivation of 
 the well known duality between D-brane
location and Chan-Paton factors, with  careful consideration 
 of the technicalities  
 regarding the translational zero modes and global holonomies,
both in the world sheet and in the target space. In order to simplify
notation we will consider a flat decoupled background $g_{0i} = b_{0i}
= 0$ and concentrate on the higher genus path integral for the field
$x^0$ which we denote by $X$ throughout this section.

\subsec{Higher genus}

 It will be
instructive to follow the opposite route to the one taken in section
2, and start with a Dirichlet path integral for a genus $g$ Riemann
surface with $N$ boundaries $B_1,..., B_N$ 
 mapped respectively to points $y_1,...,  y_N$ in the
target circle of radius $R$ (some of the $y_i$ could be identical).
The relevant amplitude is given by the periodic sum  
\eqn\periodic
{\Gamma^g (R;y_1,..., y_N) = \sum_{m_j} \Gamma^g_{\rm fixed} (R; y_j
- y_1 + 2\pi R\, m_j), }
where
\eqn\pint{
\Gamma^g_{\rm fixed}
 (R; y_1,...,y_N) = \sum_n \int_{X_{B_j} = y_j} \CD X_n \,\,
e^{-{1\over 4\pi} \int (dX_n )^2}
.}
Because of the translational invariance of the action, the path     
integral only depends on the differences of $y_j$, say $y_j - y_1$ ,    
$(j=2,..., N)$, and we may take as well $y_1 =0$. This fact was
taken into account in \periodic\ by considering only $N-1$ independent
periodic sums. In \pint,  
the $X_n$ field is multivalued around handles as    
$
\oint_{a} dX_n = 2\pi R\, n_a \in 2\pi R\,  {\bf Z}   
$, where $a= 1,...,  2g$ labels a homology basis of the Riemann
surface.

The path integral proof proceeds as usual by transforming the measure
to a first order formalism via the change of variables
$\partial_{\alpha} X \rightarrow V_{\alpha}$ (equivalently, gauging
the isometry $X\rightarrow X+ \epsilon$ and fixing the gauge $X=0$).
The transformation of the measure is
\eqn\measure{
\eqalign{
\CD X =& \prod dX \prod_{B_j} \delta( X-y_j) \cr  
=& \prod dV \prod\, ^{'} 
\delta(*dV) \prod_{a=1}^{2g} \delta\left(\oint_a V\right)
 \prod_{j=2}^{N}
\delta\left( \int_{\gamma_j} V - y_j +y_1\right) \prod_{B_j}
 \delta (V)
.}} 
Here $\gamma_j$ is a system of paths from a reference point where we
fix the translational zero mode,
 to the boundary components. It is convenient to choose the reference
 point  to lie at one particular boundary, say $B_1$. Then $X_{CM} = y_1$,     
and we have $N-1$ paths from a distinguished point on $B_1$ to the rest
of the boundary components. Once we have chosen this point, it is useful 
to drop the corresponding integration variable $X_{CM}$ at the expense   
of the delta function $\delta (X_{CM} - y_1)$. We denote this suppression 
of one integration variable by a prime in ultralocal products. In a
lattice regularization: 
$$
\CD'X = \prod_i \,^{'}  dX_i \prod_{B_j}\,^{'}  \delta (X_i - y_j)
,$$
where  $i$
denote the  points in the lattice. Following \ref\grosskleb{D.J. Gross
and I. Klebanov, \npb{344} {1990} {475.}}, the 
 complete measure \measure\ can
be motivated in the lattice as follows. The first three factors in the
second line of \measure\ account for the standard change of variables
from $X_i$ to $V_{ij} = X_i - X_j$. The flatness constraint $*dV=0$ is
represented in the lattice as $\sum_{ \partial( \rm{faces})}
 V = 0$. Since the
change of variables is linear we only have to check that we have the
same number of degrees of freedom. Prior to Dirichlet fixing we have
$N_0 -1$  
integrals over points
in the lattice,   $N_1$  link
integrals and $N_2 -1 + N +2g$ delta functions, and the matching is
ensured by Euler's theorem: 
$
N_0 - 1 = N_1 - N_2 +1 -N -2g
$. 
Now we have to express the Dirichlet deltas in first order form (in
terms of the link variables). Pick a path $\gamma$ 
in the lattice from the
reference point $X_{CM}=y_1$ to  some boundary $B\neq B_1$,  
 and label the point
variables on that boundary by $X_0, X_1,... , X_k,...$.   
Then,   the  relations
$
X_k = y_1 + V_{k,k-1} +\cdots + V_{10} + \sum_{\gamma} V 
$
can be used iteratively to prove the delta function identity:
\eqn\deltaid{ \delta\left(\sum_B V\right)
 \prod_{i\in B} \delta(X_i -y_B   ) =
 \delta\left( \sum_{\gamma} V -y_B + y_1 \right) \prod_{B}
\delta (V)
.}
For the remaining $B_1$ boundary we have
$$
\prod_{B_1}\,^{'}  \delta (X_i - y_1) = \delta(V_{10})\delta(V_{21})      
\cdots \delta(V_{mm-1})
$$
so that, together with $\delta(\sum_{B_1} V)$, we get a total factor 
of $\prod_{B_1} \delta (V)$, and we have thus succeeded in 
reducing all terms in the measure to  first order form.

 Now
we can  solve the constraints on the dual lattice
defining  $V_{ij} = V_{IJ}$ if
$(ij)^* = (IJ)$, where   $I$ denote points in the dual lattice, that is
faces of the original lattice. Then we introduce the Lagrange
multiplier field:   
$$
\prod_{{\rm faces}, I} \delta\left( \sum_{\partial I} V\right)
 = \prod_{I} \int
{d\tX_I\over 2\pi} e^{i\tX_I \sum_{J(I)} V_{IJ}}
,$$
and similarly for the rest of the terms in the measure. In continuum
notation we exponentiate the holonomy constraints as
$$
\delta\left( \oint_{a}V\right) = \int_{\bf R}\, {d\ell_a\over 2\pi} \,
e^{i\ell_a \oint_a V} = \int_{\bf R} \, {d\ell_a \over 2\pi}
 \, e^{i\ell_a \int
V\wedge h_a}
,$$
where $h_a$ is a basis for homology one-forms $\oint_a h_b =
\delta_b^a$. We can define similar one forms for each of the contours
$\gamma_j$, such that
$
\int_{\gamma_j} V = \int V \wedge h_j
$. 
They  can be written as $h_j = d\alpha_j $,  where $\alpha_j$ is an
angular variable centered at $X_{CM} = 0$, taking  values $1$ and $0$
``above" and ``below" $\gamma_j$. In particular, their integrals
along the boundaries are  
$
\oint_{B_j} h_j = \oint_{B_j} d\alpha_j = \pm 1
$, 
depending on the orientation of the contour integral. So the $h_j$ are
a basis of ``winding modes" of the open string boundaries, and  can
be used to exponentiate the global part of the Dirichlet constraints:
$$
\delta\left(\int_{\gamma_j} V - y_j +y_1\right) =
\prod_{j=2}^{N}  \int_{\bf R}\, {dp_j \over
2\pi}\,\, e^{-ip_j(y_j-y_1)} \,\, e^{ip_j \int V\wedge h_j} 
.$$
Now we can do the $V$ integral. First, we shift $V\rightarrow V- 2\pi
n_a h_a$  
 and obtain the factor  
\eqn\quantn{
\sum_{n_a} e^{-2\pi i n_a \ell_a R} = R^{-2g} \sum_{n_a} \delta \left(
\ell_a - {n_a \over R}\right) 
,}
where we have used $\int h_a \wedge h_b = \delta_{ab}$. 
This results in a quantization of $\ell_a$ in units of the dual radius
$1/R$. 
The remaining gaussian path integral to evaluate is
$$
\prod_{j=2}^{N} \int {dp_j \over 2\pi} e^{-ip_j( y_j-y_1)}
\sum_n\int_{t_{B_j}\cdot V= 0} \CD'\tX_n \,\CD V \,e^{-{1\over 4\pi} 
\int V^2 +
{i\over 2\pi} \int V\wedge (d\tX_n + 2\pi p_j h_j) }
,$$
where $\oint_a d\tX_n = {2\pi \over R} n_a$ is multivalued in the dual
circle of radius $1/R$.
The saddle point of this integral is at $V_{c\ell} = -i*(d\tX + 2\pi p_j
h_j)$. Accordingly, the Dirichlet boundary conditions $t_B \cdot
V_{c\ell} = 0$ are mapped into Neumann conditions for the fluctuating
$\tX$ field $ t_B \cdot *dX_{c\ell} = n_B \cdot d\tX_{c\ell} = 0$. 
 
 The periodic sum in \periodic\ 
has the effect of discretizing the $p_j$ via
\eqn\quantp{
\sum_{m_j} e^{-2\pi i R m_j p_j} = R^{-N+1} \, \sum_{m_j} \delta\left(
p_j - {m_j \over R}\right)
.}
In this way we  obtain the final result for T-duality of Dirichlet path
integrals at any genus:
\eqn\final{
\Gamma^g (R;y_j) = {\rm const.}\times \, R^{-2g-N+1} \sum_{m_j} \,
\delta_{\sum m} \, e^{-iy_j m_j /R} \,\, {\widetilde \Gamma}^g (
1/ R; m_1,\cdots, m_N)
,} 
where ${\widetilde \Gamma}$ is an open string path integral with
windings $m_j$ along the boundaries $B_j$:
\eqn\open{
\eqalign{{\widetilde \Gamma}^g (1/R; m_j) =&
 \sum_n \int_{n_B \cdot d\tX =0}
\CD' \tX_{nm} \, e^{-{1\over 4\pi} \int (d\tX_{nm})^2}
\cr
\oint_a d\tX_{nm} =& {2\pi \over R} n_a \,\, \,\,;\,\,\,\,\,\,\,
 \oint_{B_j}
d\tX_{nm} = {2\pi \over R} m_j
.}}
In these expressions we have introduced an extra winding number
$m_1 = - \sum_{j=2}^{N} m_j$ to obtain more symmetric looking formulas.
If all $y_j = y$ coincide, then $\Gamma^g (R,y)$ is independent of $y$
and we have the standard   partition function with the
compact volume factored out: $\Gamma (R,y) = \CZ(R,y) /2\pi R$. On the
right hand side of the duality equation \final\ we have an open string
vacuum amplitude with the zero mode also factored out. If we wish to
restore the zero mode we can write ${\widetilde \CZ}(1/R) = {2\pi \over
R} {\widetilde \Gamma} (1/R)$. Now, in terms of ${\widetilde \CZ}$ the
$R$ power counting reads
$$
\Gamma^g (R,y) \sim R^{-2g-N+2} \, {\widetilde \CZ}^g (1/R)
.$$
Since this surface comes with a coupling constant power $\sim
\lambda^{2g+N-2}$, we have the usual mapping to the dual coupling
${\tilde \lambda} = \lambda / R$.   An obvious generalization of the
preceeding remarks includes connected world sheets with several
combinations of Dirichlet and Neumann boundaries along the same
coordinate, corresponding to perpendicular D-branes in the target
space: a pair of order $(p,p')$ is T-dual to a pair $(p-1,p'+1)$.

\subsec{Non compact case}

An interesting situation is the case of a non-compact isometry, which
formally corresponds to the limit $R\rightarrow \infty$. The previous
local derivation of the duality goes through except for the 
fact that now
there are no sums over windings $n_a$ or periodic translations $m_j$.
As a result, the factors \quantn\ and \quantp\ do not appear, and
the periods of the dual field remain unquantized. This also follows from
\final\ and \open\  defining $\ell = n/R$ and $p= m/R$ and taking the
large radius limit, with the result

\eqn\noco{
\Gamma^g (\infty; y_j) = {\rm const.}\times  \,
 \int \prod_{j=1}^{N}
 dp_j \, 
 \,e^{-i p_j y_j} \,\, \delta\left( \sum p_j \right)
\,\,  
\int \, d\ell \,\, {\widetilde \Gamma}^g (\ell_a ; p_j)
,}
where ${\widetilde \Gamma}^g (\ell_a ; p_j) $ is the standard open
string path integral with the boundary conditions
$$
\oint_a d\tX_{\ell p} = 2\pi \ell_a \in {\bf R} \,\,;\,\,\,\,\,\,
\oint_{B_j}
d\tX_{\ell p} = 2\pi p_j \in {\bf R}
.$$
These  unusual boundary conditions correspond to a continuous spectrum
of winding modes, and thus represent the zero radius theory. A review
of general facts about non-compact T-duality can be found in 
Appendix B.

\subsec{Chan Paton factors}

Returning to \open, we notice that the phase factors modulating the
winding sums resemble insertions of Chan-Paton factors:
$$
e^{-iy_j m_j /R} = e^{-{iy_j \over 2\pi} \oint_{B_j} d\tX_{nm}}
.$$
In order to make the relation precise, we must symmetrize  over
the world sheet boundaries, so that each boundary can be mapped to a
particular D-brane, and we must include a combinatorial factor $1/N!$
for $N$ boundaries attached to the same brane in the target space. 
The result is equivalent to an insertion of
\eqn\cpa{    
\prod_{\rm boundaries} {\rm Tr} \, P \, {\rm exp}\left(
i\oint A_{\mu}dx^{\mu} 
\right)
,} 
where the connection is given by $A_i =0$ and $A_0 = {\rm diag} (-y_j
/2\pi)$. This corresponds to a particular $U(N_B)$ open string
background,  where $N_B$ is the maximum  
number of D-branes. In the context of unoriented strings we have to
consider $SO(2N)$ or $Sp(2N)$ groups. In this case the eigenvalues of
the connection come in pairs of opposite sign, which is compatible
with the presence of an orientifold in the D-manifold\foot{This
pairing of eigenvalues is the origin of the 16 D-branes in the dual of
type I. As we
pointed out before, the absence of a $U(1)$ factor in this case
translates into the rigid character of the orientifold.}.  

The combinatorial factor discussed above is very important, because it
serves to distinguish between several world sheet boundaries attached
to one D-brane, and the situation in which a number of D-branes lie at
the same point. In the first case we have a factor $1/N!$, whereas  we
would find $1/ N_1 ! N_2 ! \cdots $, with $N=N_1 + N_2 +\cdots$ in the
second case. Another way to notice superposition of D-branes is to
switch on gauge fields compatible with some symmetry breaking pattern
set by the connection $A_0$. Suppose that $A_0$ has a block form
$$
A_0 = -{1\over 2\pi} {\rm diag}\, (y_k {\bf 1}_{n_k})
,$$
where ${\bf 1}_{n_k}$ is the ${n_k}$-dimensional 
unit matrix. 
This  configuration 
breaks $U(N)$ to $U(n_1)\times \cdots U(n_k)\times \cdots$. 
The  most general non-abelian configuration that respects this
 symmetry
breaking pattern consists of gauge fields with an appropriate block
structure.  The corresponding Chan Paton factors take then the form
$$
\prod_{\rm boundaries} \sum_k \, e^{-{i\over 2\pi} y_k \oint dx^0} \,
\, {\rm Tr}_{U(n_k)}\, P\, {\rm exp}\left(i\oint A^{(k)}_j dx^j \right)
.$$
After T-duality we find a path integral with Dirichlet conditions at
$y_k$ and non abelian 
Chan-Paton factors $W_k$ corresponding to the $d-1$ Neumann
coordinates. By construction these factors correspond to a
world-volume $U(n_k)$ gauge theory. So we find the T-duality
implementation of Witten's observation that $n$ D-branes with
an overall 
world-volume gauge group $U(1)^n$ produce a symmetry enhancement to
$U(n)$ in the coincidence limit.  

 An interesting open problem concerns the
T-duality transformation of the most general non-abelian Chan-Paton
factor, where we may have a non trivial interaction between the $A_0$
and $A_j$ components of the connection, due to the path ordering
prescription. In view of the preceeding remarks, it is clear that this
is related in the T-dual picture 
to the interactions between proximum  D-branes (see
\ref\witten{E. Witten, hep-th/9510135.}).

\appendix{B}{Remarks on non compact T-duality }
In this appendix we recall some elementary facts about T-duality with
respect to non-compact isometries. 
\par
It should be stressed from the start that in this case the dual 
theory is not ``smooth'', at least interpreted as a sigma-model.
We shall have more to say about this towards the end of the section.
At any rate, if properly interpreted, the dual theory is 
{\it exactly} equivalent to the original one and, indeed, 
the involutive property can be explicitly checked.
\par
As stated in appendix A, the main
difference with the compact case lies in the unquantized holonomies of
the dual field in a higher genus surface.  This is a very general
consequence of Kramers-Wannier duality in the cutoff theory. Consider
for example a lattice gaussian model in a discretized Riemann surface:
$$
\Gamma^g_{\rm discrete}
  =  \prod_{i=1}^{N_0 -1} \int_{\bf R} {dX_i \over 2\pi}
\prod_{(ij)=1}^{N_1} e^{-{1\over 4\pi} (X_i - X_j )^2}  
,$$
where $N_0$ is the number of sites
 in the lattice and $N_1$ the number of links.
They are related to the number of faces by the  Euler theorem 
 $N_2 = 2-2g -N_0 + N_1$.  
Keeping track of all factors of $\pi$,  etc.,
  the same manipulations that
where introduced in appendix A  lead to  
\grosskleb: 
$$
{\widetilde\Gamma}^g_{\rm discrete} = 
\prod_{a=1}^{2g}\int_{\bf R} d\ell^a \prod_{I=1}^{N_2 -1} \int_{\bf R} 
{dX_I \over 2\pi} \prod_{(IJ)=1}^{N_1} e^{-{1\over 4\pi} (X_I - X_J + 
2\pi \ell^a \epsilon^a_{IJ})^2
}.$$
Here $I$ denotes sites in the dual lattice with $N^{*}_0 = N_2$ vertices. 
The discrete one-forms $\epsilon^a_{IJ} = \pm 1$ when the link $(IJ)$ has a
positive (negative) intersection with a homologically non trivial path in the
direct lattice, zero otherwise. There is some arbitrariness in the continuum
limit, because we can absorb ultralocal functionals into the measure. Also,
the one forms $\epsilon^a_{IJ}$ naively become delta-function distributions
localized over the path $a$. Using the freedom to redefine $X_I - X_J$
by an exact one form, it is more convenient to adopt the prescription
$$
X_I - X_J + 2\pi \ell^a \epsilon^a_{IJ} \rightarrow dX + 2\pi \ell^a h^a 
,$$
with $h^a$ a basis of harmonic
 forms with normalized intersections $\oint_a h^b =
\int h^a \wedge h^b = \delta^{ab}$ (these functions are linear in the
upper half plane). The result is 
\eqn\otraaa{
{\widetilde \Gamma}^g =
 \int_{{\bf R}^{2g}} d\ell \int \CD' X_{\ell} \, e^{-{1\over
4\pi} \int (dX_{\ell})^2},}
where $\oint_a dX_{\ell} = \oint_a (dX + 2\pi \ell_a h^a) = 2\pi
\ell_a$. In fact, we can give
 a direct proof in the continuum if we notice that
the action in \otraaa\ factorizes as  
$$
\int (dX + 2\pi \ell\cdot h)^2 = \int (dX)^2 + 4\pi^2 \ell^a G_{ab} \ell^b
,$$
where $G_{ab} = \int h_a \wedge *h_b$.
One can then explicitly perform the integral over $\ell$ with the result
$$
\int d\ell \,\, \Gamma_{\ell} =\Gamma^g\,\,
 \int d\ell \, e^{-\pi\ell^a G_{ab}\ell^b}
= {\rm det}^{-1/2} (G) \,\, \Gamma^g 
.$$
With the standard normalization for $h^a$ the matrix $G$ can be inverted by a
simplectic transformation (winding-momenta exchange),
  so the determinant is one and
we have established ${\widetilde \Gamma}^g = \Gamma^g$. 
This means that, at the level of the partition function, the
continuous holonomies are quite harmless, because they are easily
factorized. The situation is more complicated for correlation
functions, where local vertex operator insertions of momentum $q$ are
mapped into vortices of charge $q$. The dual correlator is a
frustrated partition function with local boundary conditions
 $\oint_{z_0}  
d\tX = 2\pi \,q$, where $z_0$ is the world sheet location of the operator.
 If we work in dual variables, with vortex correlation functions, the
continuous holonomies around handles are necessary to understand world
sheet factorization. One could also factorize the classical part of
the correlator, saturating the path integral with classical
trajectories $X_{c\ell}$ or  $\tX_{c\ell}$ which locally around
punctures look like
$$
X_{c\ell} (z,\zb) \sim i\,q\left( {\rm log}\,(z-z_0) +{\rm log}\,(\zb
-\zb_0 )\right)
$$
$$
\tX_{c\ell} (z,\zb) \sim i\,q\left( {\rm log}\,(z-z_0) - {\rm
log}\,(\zb-\zb_0) \right).
$$  
 
We can also arrive at the same result by explicitly taking the
$R\rightarrow \infty$ limit of the compact case.
 The partition function
 at genus $g$ is given by 
$$ \CZ^g (R) = \sum_{n\in {\bf Z}^{2g}}\,\, \CZ^g_n (R) $$ $$
\CZ^g_n (R) = \int\CD X_n\,\, e^{-{1\over 4\pi} \int (dX_n )^2}
,$$  
where $\oint_{a} dX_n = 2\pi R n_a$.  
Fixing the translational zero mode: 
$\int\CD X = 2\pi R \int \CD'X$ we can  
 define $\Gamma^g (R) = \CZ^g (R) /
2\pi R$.
Now, T-duality is the statement
 $$
\Gamma^g (R) = {\rm const.}\,\,\times  R^{-2g}\,\, \Gamma^g (1/R)
.$$
In the large radius limit no windings survive, so
$$
\lim_{R\rightarrow 0} \Gamma^g (1/R) = \Gamma^g_0 (\infty) = 
\lim_{R\rightarrow 0}\, \sum_n \,\, R^{2g}\, \Gamma^g_n (R)
.$$
Finally, defining  $\ell = n R \in {\bf R}^{2g}$, and taking the
limit, we arrive at \otraaa:       
$$
\Gamma^g_0 (\infty) = \int_{{\bf R}^{2g}}\, d\ell\,\, \Gamma^g_{\ell}
.$$

So, our non-compact dual is the same as the zero radius limit, with
the dilaton shift reabsorbed (note that the rescaling factor $R^{-2g}$
was needed to transform the discrete winding sum into a continuous
integral). The occurrence of the continuous holonomies $\ell \in {\bf
R}^{2g}$ can be interpreted in two ways. We can say that we do not
have a sigma model any more, because the embeddings are discontinuous
in terms of the $\tX$ field living in $\bf R$. We can also interpret the
holonomies as a continuum of twisted sectors. After all, the windings
build the twisted sector of the orbifold ${\bf R} /2\pi R {\bf Z}$ 
and, as $R$ goes to zero, the winding spectrum becomes continuous.
Furthermore, the invariant part of the spectrum (the momentum modes)
dissappears. So we can regard the non-compact dual as a sigma model
in the ``orbifold" \ref\rattickw{J.J. Atick and E. Witten,
\npb{310} {1988} {291.}} 
$$   
\lim_{R\rightarrow 0} {{\bf R}\over 2\pi R {\bf Z}} \equiv {{\bf
R}\over 0}
.$$

\appendix{C}{Canonical Transformations in the operator formalism}
When dealing with closed strings, it is well-known that T-duality can be interpreted as a canonical transformation \ref\can{E. Alvarez, 
L. Alvarez-Gaum\'e and Y. Lozano, \plb{336} {1994} {183.}}.
Although for open strings the dual theory is somewhat exotic, it is still true that it can be formally interpreted as a canonical 
transformation in a big Hilbert space.
\par

In the hamiltonian formalism T-duality corresponds to the following
 canonical
transformation in the Hilbert space on the circle:
$$
{\widetilde \Psi}(\tX) = \int \CD X \, e^{-iF(X,\tX)} \Psi (X)
,$$
where $\Psi$ and ${\widetilde \Psi}$ are wave functionals and the 
integral is over a basis of the configuration space on the circle (not
a path integral). The generating functional is very simple:
$$
F(X, \tX) = {1\over 2\pi} \oint_{S^1} \tX\, dX
.$$
In this appendix we  explicitly construct the generating functional
that implements T-duality for the boundary states of Neumann or
Dirichlet character. The most general such  
 boundary state (in the matter sector) was constructed in ref.
 \clny
$$
|B\ket_{\eta} = \int \CD \bx\, \CD x\, \CD \btheta\,
 \CD \theta \,\, e^{-S_B}\, \,
|\bx, x\ket \otimes |\btheta, \theta\ket_{\eta} 
,$$
where $\eta = \pm$ labels the spin structure in the open string
channel  and, after T-duality, distinguishes branes from anti-branes.
 The configuration space  coordinates  satisfy ($m>0$)
$$
(a_m^{\mu\dagger} + {\overline a}_m^{\mu} - \bx_m^{\mu} )|x,\bx\ket =
0
$$
$$
(a_m^{\mu} + {\overline a}_m^{\mu\dagger} - x_m^{\mu})|x,\bx\ket = 0
$$
$$
(\theta_m^{\mu} - \psi_m^{\mu} + i\eta\, \bpsi_m^{\mu\dagger} )
|\theta,\btheta\ket_{\eta} =0
$$
$$
(\btheta_m^{\mu} - \psi_m^{\mu\dagger} -i\eta \,\bpsi_m^{\mu}
)|\theta,\btheta\ket_{\eta} = 0
,$$
where the index $m$ in $\psi_m$ runs over the integers in the R-R
sector of the closed string, and half integers in the NS-NS sector.
  The relation of
$a_m^{\mu}$ with the standard notation of bosonic oscillators is
$\alpha_m^{\mu} = -i\sqrt{m}\, a_m^{\mu}$. The normalized ``position"
eigenstates are 
$$
|x,\bx\ket = {\rm exp}\left( -{1\over 2} \bx \cdot x - a^{\dagger}\cdot
\ba^{\dagger} + a^{\dagger} \cdot x + \bx \cdot \ba^{\dagger} \right)
|0\ket
$$
$$
|\theta,\btheta\ket_{\eta} = {\rm exp}\left(-{1\over 2}
\btheta\cdot\theta +i\eta \,\psi^{\dagger}\cdot\bpsi^{\dagger}
+\psi^{\dagger} \cdot \theta -i\eta \,\btheta\cdot
\bpsi^{\dagger}\right) |0\ket
.$$
The dot product is defined by
$$
\bx\cdot x = \sum_{\mu=1}^{D} \sum_{m>0} \bx_m^{\mu}
x_m^{\mu}
,$$
and the position states are conveniently normalized to one, so that
the free (Neumann) boundary states corresponding to $S_B =0$ are
$$
|{\rm Neumann}\ket_{\eta} = {\rm exp}\left( a^{\dagger}\cdot
\ba^{\dagger} -i\eta \,\psi^{\dagger}\cdot \bpsi^{\dagger}\right) |0\ket
.$$
The zero modes in the R-R sector carry a representation of the
$SO(1,9)$ Clifford algebra, and the vacuum satisfies
$$
(\psi_0^{\mu} -i\eta\, \bpsi_0^{\mu} ) |0\ket_{\eta} = 0
.$$

Under T-duality we have $\ba^{\dagger} \rightarrow -\ba^{\dagger}$ and
$\bpsi^{\dagger} \rightarrow -\bpsi^{\dagger}$ (for simplicity of
notation, we consider here the duality transformation of all
coordinates, so that we are dealing with the D-instanton).
 We know that this is
generated by a canonical transformation with generating function of
the form $F(q,{\tilde q})$, where ${\tilde q}$
 are the dual variables. It is very easy
to find such  function by simple gaussian integral
manipulations requiring
$$
|{\tilde q}\ket_{\rm dual} = \int \CD q\,\, e^{-iF({\tilde q},q)} |q\ket
.$$
The sign of the anti-holomorphic components is inverted with the
following generating function:  
$$
F(x,\bx,\theta,\btheta;\tx,{\tilde {\bx}},{\tilde {\theta}},
{\tilde {\btheta}}) = {i\over 2}\left(
\bx\cdot \tx - {\tilde {\bx}}
 \cdot x + \btheta \cdot {\tilde {\theta}} - {\tilde {\btheta}} \cdot
\theta \right) + {\rm constant}
.$$
 Introducing mode expansions
$$
X(\sigma) = q + {1\over \sqrt{2}} \sum_{m>0} {1\over \sqrt{m}} \left(
x_m \,e^{-im\sigma} + \bx_m \, e^{im\sigma} \right)
$$
$$
(\psi + i\eta\, \bpsi)(\sigma) = \psi_{\eta} (\sigma) = 
 {1\over \sqrt{2}}\sum_{m>0} \left(
\theta_m \, e^{-im\sigma} + \btheta_m \, e^{im\sigma} \right)
,$$
and similarly for the dual fields, we obtain the final result
$$
F(X,\tX;\psi_{\eta}, {\widetilde \psi}_{\eta}) = {1\over 2\pi} \oint_{S^1}
\left(\tX \, dX - i{\widetilde \psi}_{\eta} \, \psi_{\eta} \right)
,$$
where the fermionic terms are restricted to the non-zero modes. In
general, the zero mode part  of the dual state is determined by the
constraint:  
$$
({\widetilde{\psi}}_0 +i\eta \,{\widetilde{ \bpsi}}_0 )^{\rm tr}
 |0\ket_{\eta} =0
,$$
where ``tr" stands for the transverse coordinates to the $p-$brane
(all the coordinates for the case of the D-instanton).
An analogous generating function can be written in the super-ghost
sector.

\listrefs
\bye